# k-Means Clustering in Fingerprint-Based Configuration Selection for Fitting Interatomic Potentials

Miroslav Lebeda[a, b], Jan Drahokoupil[a, c], Ludvík Löbel[c], Petr Vlčák[a]

[a] *Department of Physics, Faculty of Mechanical Engineering, Czech Technical University in Prague, Technická 4, 16607 Prague 6, Czech Republic*

[b] *Department of Solid State Engineering, Faculty of Nuclear Sciences and Physical Engineering, Czech Technical University in Prague, Trojanova 339/13, 12000 Prague 2, Czech Republic*

[c] *Department of Material Analysis, FZU – Institute of Physics of the Czech Academy of Sciences, Na Slovance 1999/2, 18221 Prague 8, Czech Republic*

**Corresponding author:** Miroslav Lebeda, lebedmi2@cvut.cz

## Abstract

In this study, we present a method for selecting an arbitrary number of distinct configurations from a larger dataset using atomistic configuration fingerprints and k-means clustering. This approach improves the accuracy of fitting classical molecular dynamics interatomic potentials to DFT data for both energies and forces, while requiring fewer configurations than random sampling. We demonstrate this improvement by fitting an embedded-atom method (EAM) potential for titanium, using various configurational sizes from an initial set of 1800 configurations. The k-means clustering consistently achieves better precision and lower standard deviations for a smaller number of configurations than random selection. The results also suggest that only about 30 configurations are sufficient to obtain an EAM model that well describes the full set of 1800 configurations in terms of energies and forces. Additionally, t-SNE was used to reduce the configuration fingerprints into 2D space and it revealed an overlap between two configuration subsets with and without Ti vacancy, indicating similar atomic environments. This similarity is captured by k-means clustering but not by random selection. Furthermore, when configurations with vacancies were excluded from the k-means algorithm and used only as a test set, their energy and force predictions showed similar precision to when they were included. This indicates that the overlapping configurations in the 2D t-SNE space indeed imply potential information redundancy among the atomistic configurations.

# 1. Introduction

In the development of classical or machine-learning interatomic potentials based on *ab initio* density functional theory (DFT) data, an important task is the selection of suitable atomistic configurations for fitting or learning [1–3]. This selection should aim to minimize the training dataset size while maximizing the representation of the underlying potential energy surface (PES). A smaller yet well-selected dataset offers several advantages: (1) it reduces the time and computational resources required to fit potential parameters or train neural networks, which are often computationally intensive; (2) it helps prevent overfitting, thereby improving the potential's ability to better generalize to new, unseen configurations; and (3) it minimizes the time and computational expense associated with also underlying calculations of DFT data. Although machine learning models typically benefit from larger datasets, selecting non-redundant configurations reduces computational demand in terms of both data generation and model training, without sacrificing learning quality.

The configurational selection task has recently become an active area of research[4]. For classical potentials, configurations are typically chosen uniformly over time or by uniform random atomic displacements and deformations using additional software such as Atomsk[5]. For example, in[6], over 7000 configurations were prepared by taking equidistant snapshots from the last 7.5 ps of *ab initio* molecular dynamics (AIMD) run with initial potential. However, this method can result in certain configurations with similar local atomic environments, possibly leading to redundancy in the data used for fitting. An advantageous alternative could be to generate a larger number of configurations and then select the most distinct 7000 configurations to ensure a diverse and informative dataset.

For the selection of the most distinct configurations, techniques based on atomistic descriptors are often used[7,8]. A descriptor in this context is a numerical representation of molecules, molecular clusters, or material structures that remain invariant under translations, mirroring, rotations, and atomic permutations operations. For instance, one of the used descriptors is the Smooth Overlap of Atomic Positions (SOAP)[9], which characterizes the local environment of each atom within a configuration by considering the spatial distribution of its neighboring atoms. Specifically, in SOAP, the atomic density around each atom is modeled as a sum of Gaussian functions centered on its neighboring atoms. The overlap integral of this atomic density with those of the neighboring atoms from the same configuration is calculated, providing a numerical description of the local environment information of each atom within the configuration.

Another valuable descriptor, although only applicable on periodic systems without vacuum, is a CrystalNN (crystal nearest neighbors' algorithm)[10]. The CrystalNN describes coordination numbers and evaluates similarity to specific atomic geometries. This makes it particularly useful for providing an easily interpretable understanding of significant local geometries within a dataset of structural fingerprints. In contrast, methods like SOAP also capture local geometries but rely on abstract Gaussian expansions, which are less easily interpretable than CrystalNN's components. Compared to methods like the radial distribution function (RDF), which describes statistical information about distances between pairs of atoms in the system, CrystalNN focuses on geometrical relationships between atoms. However, CrystalNN does not distinguish between different lattice parameters, meaning that structures with identical atomic arrangements but differing atomic distances would be deemed the same. By combining CrystalNN with RDF, both local geometrical information and atomic distance data can be captured. A good overview of atomistic descriptors and their implementation within a Python library is available here[8,11]. The atomistic descriptors can then be used to compare different configurations with each other or to generate a diverse set of configurations using clustering methods such as k-means.

Besides the clustering selection methods, there exists active learning algorithms that iteratively select the most informative configurations as they venture into poorly defined areas of PES[12]. This approach improves model accuracy by targeting uncertain regions, but it typically requires retraining the model multiple times throughout the process, potentially making it computationally expensive. Additionally, active learning may sometimes put too much focus on uncertain areas while neglecting the broader diversity of configurations which may also be important for the potential model. In contrast, the clustering methods based on descriptors do not require retraining, as no new configurations are added during the fitting process. These methods ensure a diverse and non-redundant set of configurations, but since they rely solely on structural similarity, they may miss less distinct configurations that could be important for accurately capturing specific material properties. It is also possible to combine both approaches. For example, the clustering could first be used to ensure a diverse set of atomic environments, and then active learning could be applied to refine the model by focusing on uncertain regions.

Despite the success of using descriptors in recent times, variability in their definition and their effectiveness remains still a question in research[13]. Connected to that, there is a need for user-friendly, automated selection tools to reduce the number of configurations used for the DFT calculations and fitting. These tools should streamline the process by selecting the most relevant configurations (in order to remove redundancy due to similarity of atomic environments) from a larger pre-generated dataset, such as those obtained through uniform time sampling.

In this work, we present a technique for selecting an arbitrary number of the most distinctive configurations from an existing set. Our method characterizes the distinctiveness of local atomic environments using a combination of the following fingerprints: (1) CrystalNN (crystal nearest neighbors' algorithm, for differences in local atomic geometry), (2) atomic mass difference (for differences in chemical composition), (3) average bond distance and radial distribution function (for differences in distances between atoms and thus also in lattice parameters). Each configuration is assigned such unique fingerprint, and these fingerprints are then subjected to dimensionality reduction via Principal Component Analysis (PCA)[14] before being clustered into groups using the k-means algorithm. The number of clusters is set to the user-defined number of configurations to select. From each cluster, the configuration closest to the centroid is chosen to ensure maximum variability in the configuration set. In order to increase interpretability and to possibly observe underlying patterns in the configurations, the dimensionality of the fingerprints is further reduced to 2D using t-distributed Stochastic Neighbor Embedding (t-SNE) method[15].

We chose the CrystalNN method, combined with RDF and statistical information on bond distances, to evaluate its suitability as a descriptor. Our goal was to determine whether this combination could effectively facilitate k-means clustering to identify distinct structures suitable for fitting interatomic potentials, specifically in terms of the accuracy of energy and force predictions. To the best of our knowledge, no prior studies have conducted such an assessment using CrystalNN as a descriptor in conjunction with atomic distance-based descriptors.

The advantage of the k-means clustering-based selection is demonstrated through comparison with uniformly randomly chosen configurations for the fitting of a titanium (Ti) classical embedded atom method (EAM) potential on DFT data (total energies and atomic forces). To illustrate this, 1800 atomistic Ti configurations were generated. The primary goal was not to develop a Ti potential that well describes its material properties but to statistically showcase the effectiveness of the k-means clustering algorithm regarding the fit quality of energies and forces. The results consistently show smaller errors in energies and forces for smaller number of configurations when they are selected using our algorithm rather than by uniformly random selection. Additionally, EAM fitted on configurations selected by clustering are consistently more reliable, as evidenced by lower standard deviations in fit quality of energies are forces. The t-SNE projection into 2D space revealed distinct 5 groups and an overlap between configuration groups with and without Ti vacancies, suggesting similar atomic environments. This overlap was captured by K-means clustering but not by random selection. The presented code for calculating fingerprints and selecting user-defined number of configurations via k-means clustering is available here github.com/bracerino/Selecting-distinct-set-of-structures with a full description.

# 2. Methodology

## 2.1 Configuration Fingerprint and K-means Clustering

The fingerprint characterizing each atomistic configuration was defined as consisting of three parts:

1) **CrystalNN** (244 elements, for geometrical atomic environments)

   The CrystalNN method generates a 61-dimensional vector for each atom within a configuration, capturing various aspects of its local atomic coordination environment using coordination descriptors. These descriptors quantify the likelihood of different coordination numbers and the resemblance to specific geometric configurations. For instance, they describe the likelihood of 2-fold coordination, the resemblance to an L-shaped coordination geometry, to the geometry of a water molecule, to the geometry where two atoms form a bond angle of 120 ° with the central site atom, etc. The values of these descriptors range from 0 to 1, where 0 indicates no resemblance to the target motif while 1 indicates a perfect match. Following the generation of atomic site-specific vectors, statistical measures are performed across them for all atoms within a configuration. This involves the mean, standard deviation, minimum and maximum values, resulting in 244-dimensional CrystalNN fingerprint (61 elements * 4 statistical measures). The CrystalNN method is specifically designed for periodic structures that do not contain vacuum. Its application to molecules, clusters, and surfaces typically leads to computational errors. Thus, in cases where CrystalNN is not applicable, all elements of the CrystalNN vector are assigned a zero value. The differentiation between these types of structures is then reliant on the other parts of the total fingerprint. However, this would decrease accuracy as no local atomic geometries would be present in the fingerprints. If configurations with vacuums need to be clustered and reduced based on structural redundancy, an alternative descriptor like SOAP could be used instead of CrystalNN part.

2) **Atomic mass difference** (4 elements, for differences in local element composition)
   This part represents the aggregate differences in atomic masses between an atom and its neighbors (up to 10 Å) in the configuration. The values are averaged over each atomic site, resulting in four metrics: mean, standard, minimum, and maximum value.

3) **Average bond distance** (4 elements), **radial distribution function** (100 elements, for different distances between atoms / lattice parameters)
   The first 4 elements represent statistical measures (mean, standard deviation, minimum and maximum) of the average bond distances for each atomic site and its neighboring pairs (up to 10 Å) within a configuration. The other 100 elements represent the radial distribution function (RDF), which is calculated and averaged across all atomic sites, with a distance cutoff of 10 Å, a spacing of 0.1 Å between points, and Gaussian smoothing with a width of 0.1 Å. This results in 100 elements for the RDF. The choice of 100 points was determined through internal testing to balance the ability of the algorithm to distinguish configurations that differ only by atomic distances (structures which are uniformly deformed). This number is, however, arbitrary, and it can be adjusted as needed.

After computing the fingerprints for the complete set of configurations, data preprocessing was performed using standardization, so that each feature had a mean of 0 and a standard deviation of 1. Subsequently, the weight for the atomic mass part of the fingerprint was increased by a factor of 4 (although this adjustment is not reflected for the presented fitting of Ti as there is only one element). The weight for the average bond distance component was increased by a factor of 2. These adjustments were based on internal considerations and testing, and can be possibly modified to better suit other applications.

To improve performance and focus on the most important features of the standardized fingerprints, we applied Principal Component Analysis (PCA). We then selected the PCA components that accounted for up to 98 % of the total variance in the data, reducing the dimensionality of the original 352-dimensional fingerprints. These reduced PCA components were used as new descriptors for each configuration and then clustered using the k-means algorithm. Thus, the k-means clustering was performed on the reduced-dimensional PCA components, not the original fingerprint data. This ensured that configurations with similar structural characteristics, as defined by the reduced PCA descriptors, were grouped together in the same cluster. Such a clustering process is an unsupervised learning method, as the configurations were grouped based solely on their structural fingerprints, without relying on any labels or predefined categories.

The number of clusters was set to match the number of configurations which were to be selected as the most distinct ones. To identify the most distinct configurations, a practical strategy involves selecting a representative configuration from each cluster that is closest to the cluster's centroid. This approach ensures that the most unique configurations are identified (within this algorithm). Due to the inherent randomness in the initial selection of cluster centroids in k-means, the representative configurations may vary for each run, nonetheless this method consistently yields a representative set of configurations. Specific k-means settings that we used included a tolerance of $1 \times 10^{-6}$, Lloyed's algorithm, and a maximum of 600 iterations.

It should be noted that, in some cases, very similar atomic environments may be important for capturing specific material properties. These environments may not be distinguished and thus selected by our method, as it focuses on deleting redundant configurations based on structural similarity to create a reasonably good frame for the interatomic potentials while minimizing computational costs. However, when such similar environments are important, they should ideally be identified *a priori* or addressed with methods that go beyond structural similarity. These cases can then be included alongside the broader set of configurations selected by the presented method.

We further employed t-distributed Stochastic Neighbor Embedding (t-SNE) with a perplexity parameter of 30 to reduce the still possibly high-dimensional PCA-reduced fingerprints into a 2D space, creating human-readable visualizations and potentially revealing underlying patterns between the configurations. This technique prioritizes preserving local distances between fingerprints, meaning that points which are close to each other in the original high-dimensional space should remain close in the reduced 2D space.

## 2.2 Generation of Ti Atomistic Configurations

A 3×3×2 Ti supercell (36 atoms) with initial lattice parameters of a conventional cell $a$ = 2.950 Å, $c$ = 4.686 Å ($c$ / $a$ = 1.588)[16] was constructed and subjected to DFT geometry optimization. By applying Atomsk software on this optimized structure, 300 structures (configurations) were generated by randomly displacing atoms in the x, y, and z directions with a maximum displacement of 0.1 Å, another 300 with a maximum displacement of 0.2 Å, and another 300 with a maximum displacement of 0.5 Å. An example configuration from each category is shown in **Fig. 1**. Each of these 300 structures was further deformed by applying strain from <-2, +2) % range equidistantly in the x, y, and z directions. This way, total of 900 configurations were generated. A second supercell with ~3 % of Ti vacancies was created by deleting one atom from the initial optimized Ti supercell. The same set of random displacements and deformations was applied, generating an additional 900 configurations.

In total, 1800 configurations were prepared. From this set, we were selecting configurations for fitting the EAM potential in quantities of 5, 10, 15, 20, 25, 30, 40, 50, 60, 70, 80, 90, and 100. These subsets were chosen using the k-means clustering algorithm, where the number of clusters was set equal to the number of configurations in each subset, and the configuration closest to each cluster centroid was chosen. For comparison, we also performed uniform random selection of configurations for the same subset sizes. The computational time for calculating the fingerprints to be clustered using the proposed method was under 1 hour for the full set of 1800 configurations on a single processor core within the Python code. This time could be reduced by parallelizing the fingerprint calculations across multiple cores. Importantly, the fingerprints only need to be calculated once and can potentially be reused later on when needed. The PCA analysis and clustering itself took only a few seconds to select, for example, 100 configurations. The t-SNE projection from the PCA-reduced fingerprints into 2D space took less than 1 minute. Given such low computational demand, it is efficient to perform configuration selection based on the k-mean algorithm prior to DFT calculations.

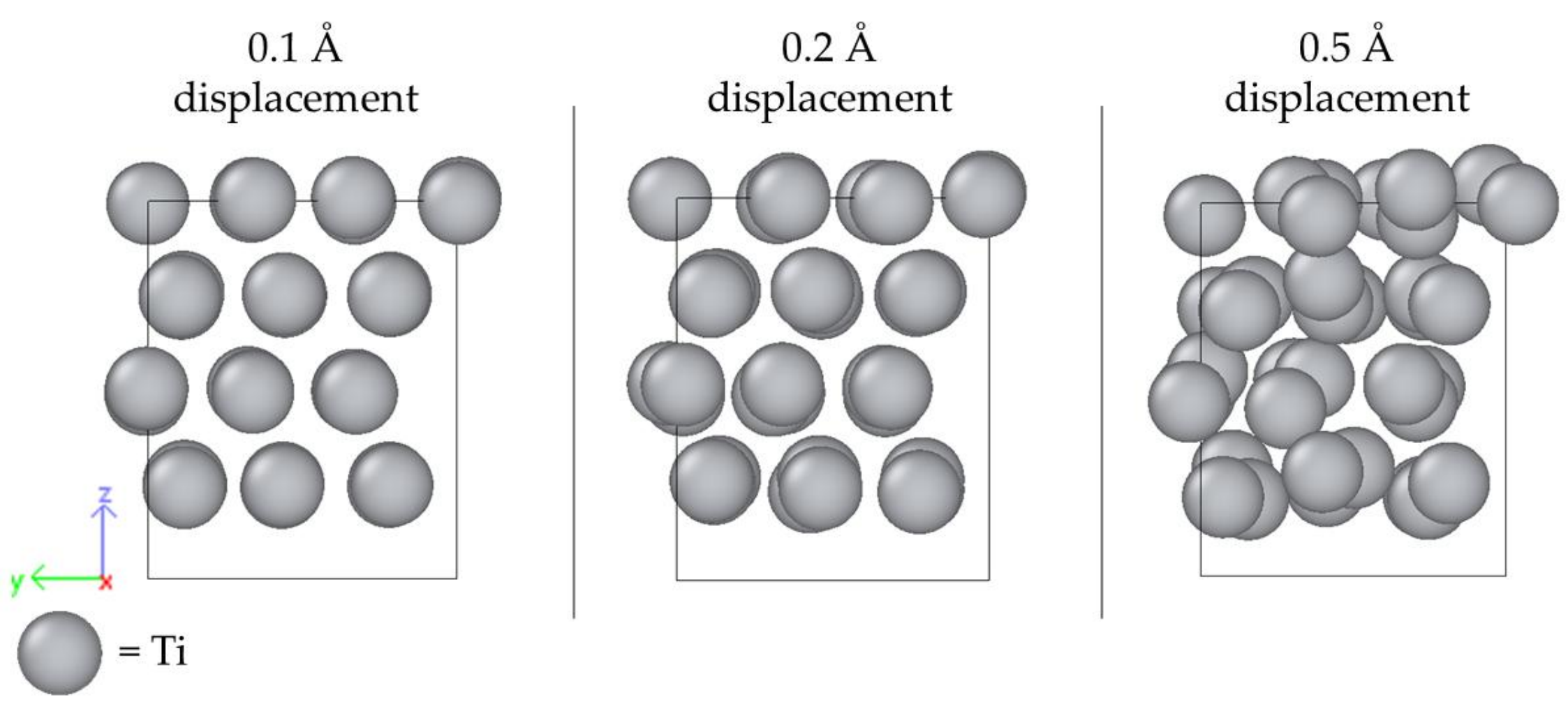

**Fig. 1:** Illustration of three generated configurations from the initial Ti supercell with 36 atoms, showing maximum displacements of 0.1 Å, 0.2 Å, and 0.5 Å, respectively.

## 2.3 DFT and MEAMfit2 Settings

For each configuration, we performed single-point DFT calculations to obtain the total energy and the forces acting on the atoms. These calculations were carried out using the Vienna Ab initio Simulation Package (VASP)[17,18]. The Generalized Gradient Approximation (GGA) functional with Perdew-Burke-Ernzerhof (PBE) parametrization was employed. Energy convergence tests on the energy cutoff and k-point sampling were performed, and their values were set to achieve convergence within 1 meV as follows: the energy cutoff of 625 eV and a Monkhorst-Pack k-point grid of 6 × 6 × 5 (for the Ti supercells). The minimum distance between atoms in each of the 1800 configurations was determined using the radial distribution function (RDF), revealing a minimum distance of 1.6 Å. This ensured that no interatomic distance was smaller than the Wigner-Seitz radius parameter of 1.323 Å for Ti within VASP. The ground-state energy for each of the 1800 generated configurations was calculated.

The fitting of the EAM potential to the DFT data obtained from VASP was conducted using MEAMfit2[19], with a cutoff distance of 4.4 Å, and the number of terms in the pair potential, electron densities, and embedding function all set to 3. The general fitting workflow in MEAMfit2 begins by generating initial potentials with random parameters, optimizing them using the conjugate gradient method, and selecting the best ones. The next stage involves further refinement of the top potentials using a genetic algorithm. In this process, two potentials are randomly selected, combined, and occasionally mutated, followed by optimization with the conjugate gradient method. If the resulting potential has a lower optimization function than any of the current top potentials, it replaces the least optimal one in the list. This process repeats until either a convergence criterion is met or the specified time limit is reached. During the genetic algorithm fitting phase of MEAMfit2, we utilized a list of 50 top potentials. All other settings were kept at their preselected values within the software.

For each configuration subset size (5, 10, 15, ..., 100 configurations) used for fitting, we selected the corresponding number of configurations 50 times using uniform random selection and 50 times using the k-means clustering algorithm. This resulted in 50 configuration groups for each subset size, both for random selection and k-means clustering. Each group was used to fit the EAM potential, allowing us to generate statistics on the precision of the fitted potentials for each corresponding subset size, measured by the root-mean-square error of energies and forces compared to the DFT data. Therefore, for each subset size and for each selection method, we obtained 50 fitted potentials, from which we calculated the median root-mean-square error (RMSE) of energies and forces, as well as their standard deviations. The entire set of 1800 DFT data, including those used for fitting, was employed as the test dataset to evaluate the obtained potentials.

# 3. Results

## 3.1 Clustering of Ti Configurations

The 352-element fingerprints of the generated Ti set, consisting of 1800 configurations, were analyzed using PCA. The first 85 components accounted for a cumulative explained variance of 98% (**Fig. 2**). This number of components was used for each configuration as its reduced fingerprint, and they were clustered using k-means into the chosen size of the subset for fitting.

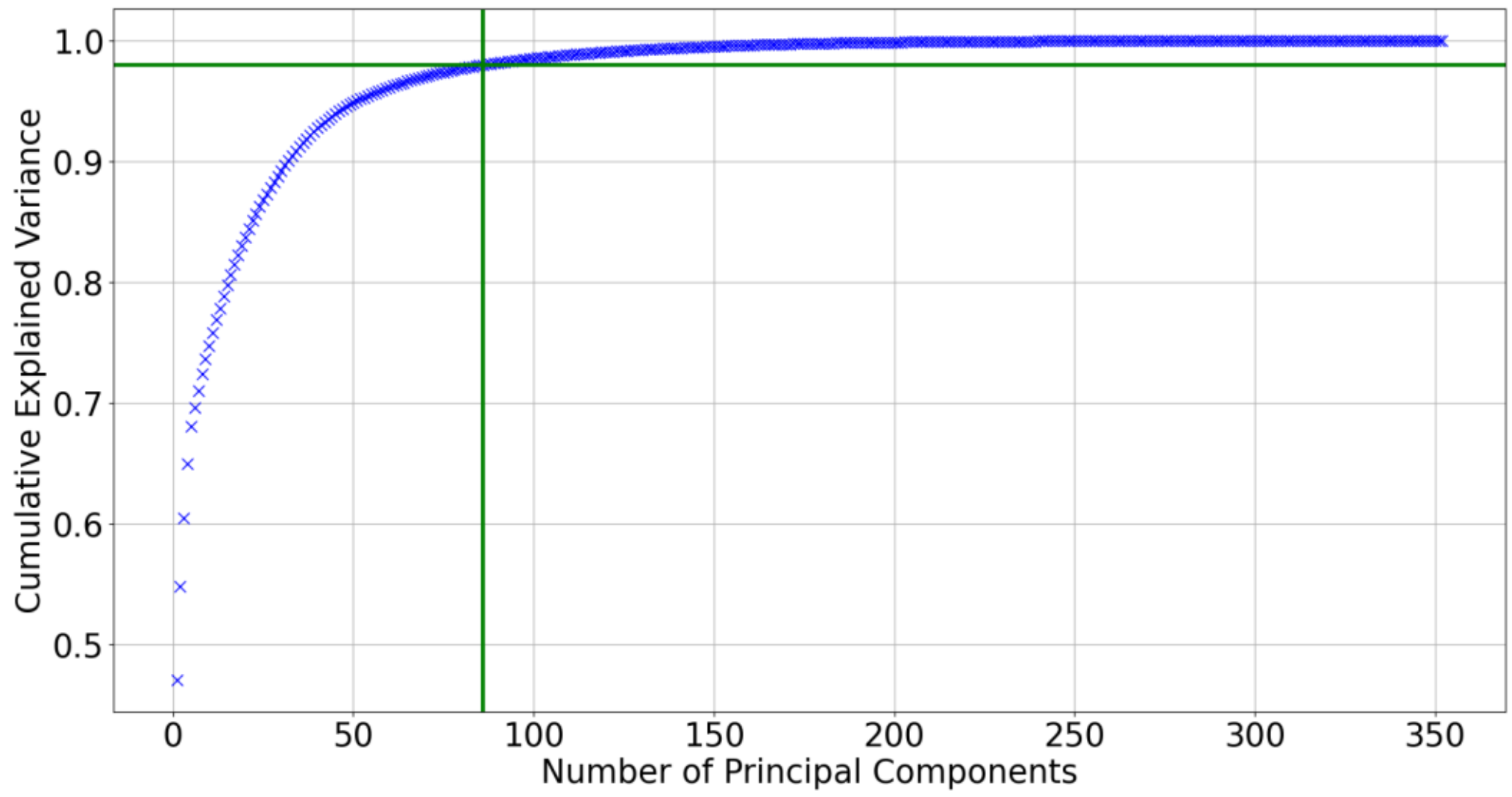

**Fig. 2:** Principal Component Analysis (PCA) of 352-element fingerprints for a dataset of 1800 titanium atomistic configurations. The cumulative explained variance reached 98 % with the first 85 components, as indicated by the green lines. These 85 components were then used as a new, reduced-dimensional fingerprint for each configuration.

The significance of each of the 352 initial features of the fingerprints is depicted in **Fig. 3**, represented by the sum of squared loadings for each feature across the 85 PCA components. Naturally, the atomic mass difference has zero contribution, since only one element (Ti) is present in the dataset. All other parts of the configuration fingerprints have an important contribution to the 98 % explained variance. This can be linked to the structural properties in terms of the fingerprint variances across the 1800 Ti configurations.

The first part of the original configuration fingerprint (components 0 to 243) corresponds to CrystalNN metrics, which are related to coordination numbers and their local atomic environment geometries. Specifically, we can associate components with higher loadings (greater e.g. than 0.6) to the structural features that show the most variance in the 1800 Ti configuration dataset. Note that the higher loading for the component means that this component has a larger variance in the fingerprints of the configuration dataset, i.e., its value in the fingerprint dataset is more diverse. The components with loadings above 0.6 correspond to specific structural features as follows (first number is the index going from 0 to 243):

10 – minimum value of coordination number 2; 14 = minimum value of the L-shaped 2-fold geometry; 18 = minimum value of the water-like 2-fold geometry; 34 = minimum value of coordination number 3; 38 = minimum value of the trigonal planar 3-fold geometry; 42 = minimum value of the trigonal non-coplanar 3-fold geometry; 46 = minimum value of the T-shaped 3-fold geometry; 58 = minimum value of the tetrahedral 4-fold geometry; 62 = minimum value of the rectangular see-saw-like 4-fold geometry;

120 = mean value of the body-centered cubic 8-fold geometry, 122 = minimum value of the body-centered cubic 8-fold geometry; 123 = maximum value of the body-centered cubic 8-fold geometry; 187 = maximum value of the q2 12-fold geometry; 191 = maximum value of the q4 12-fold geometry; 198 = minimum of coordination number 13.

The second non-zero part of the fingerprint (components 248–251) relates to bond distance statistics, which reveal considerable variance in atomic pair distances within the 1800 Ti configurations. This suggests that bond lengths are an important distinguishing factor in the fingerprint dataset. The last part of the fingerprint (252 – 351 components) corresponds to RDF, covering atomic pair distances from 0 Å to 10 Å in 0.1 Å intervals. The RDF loadings indicate at which specific distances are the largest variances between the RDFs of all 1800 configurations. For instance, the peak around 1.9 Å marks the distance with the large variation among the Ti configurations.

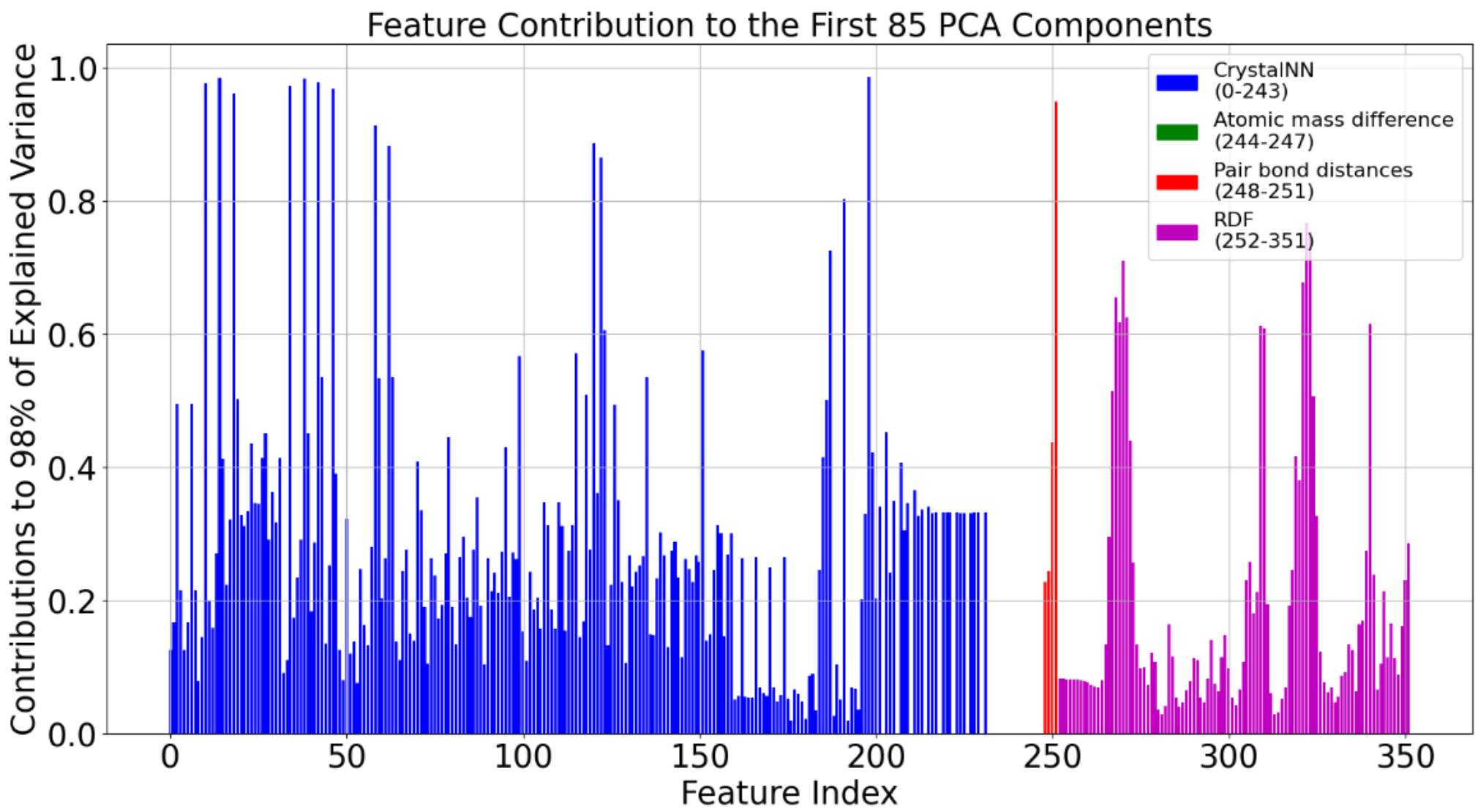

**Fig. 3:** Contributions of each of the 352 fingerprint features to the 98 % of cumulative explained variance (corresponding to the first 85 PCA components) within the 1800 titanium set of configurations. The contributions are calculated by summing the squared loadings of each feature across all 85 PCA components. The green part corresponding to the atomic mass difference is zero as there is only one element in the dataset.

To convert the 85-dimensional fingerprints into a human-readable format and to visualize potential patterns between the 1800 Ti configurations, we employed t-SNE dimensionality reduction into a 2D space (**Fig. 4**). This visualization revealed five distinct groups:

1. One group includes the supercells without vacancies and with a maximum atomic displacement of up to 0.1 Å (marked as 'Standard 0.1').
2. The second group consists of supercells with no vacancy and displacements of up to 0.2 Å ('Standard 0.2').

3. The third group features supercells with one Ti vacancy and a maximum displacement of 0.1 Å ('Vaca 0.1').
4. The fourth group contains supercells with one Ti vacancy and a maximum displacement of 0.2 Å ('Vaca 0.2').
5. The fifth group is formed by supercells without vacancies and with a maximum displacement of 0.5 Å ('Standard 0.5'), as well as supercells with one Ti vacancy and a maximum displacement of 0.5 Å ('Vaca 0.5').

The fifth group exhibits considerable overlap between non-vacancy and vacancy Ti supercells, indicating redundancy (i.e., similarity in atomic environments) between these two subsets. This potential redundancy cannot be taken into account when selecting configurations by uniform random selection, but it is effectively identified through k-means clustering, highlighting the advantage of using this algorithm. To assess the potential similarity in energies and forces between the 'Standard 0.5' and 'Vaca 0.5' subsets, the 'Vaca 0.5' subset was removed from the configuration selection and described using the remaining configurations. The results of this analysis can be found in Section 3.3.

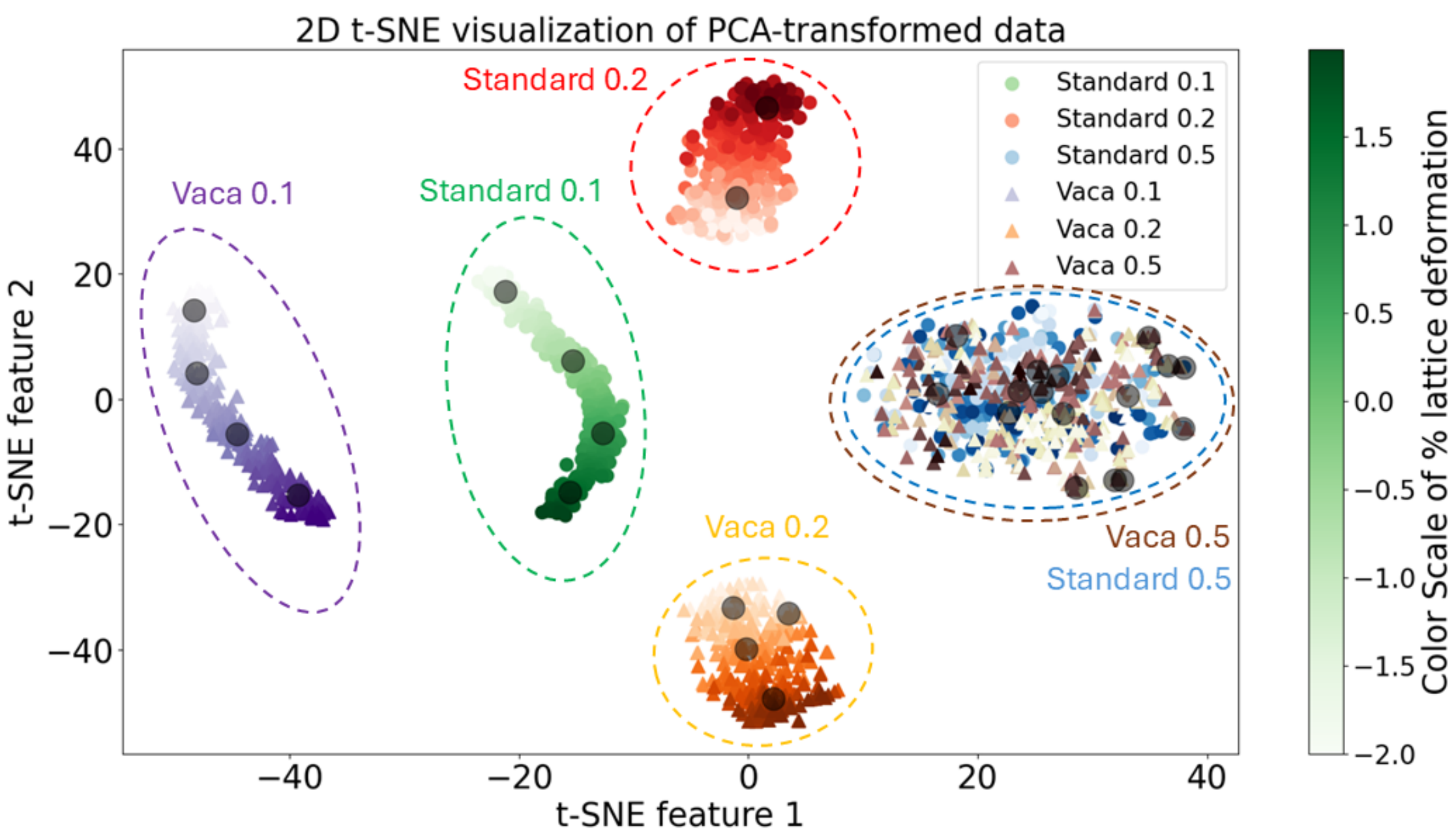

**Fig. 4:** Conversion of 85-element fingerprints for a dataset of 1800 titanium atomistic configurations into 2D space using t-distributed Stochastic Neighbor Embedding (t-SNE). The color maps indicate percentage lattice deformations from -2 % (lighter color) to 2 % (darker color). 'Standard' refers to the Ti supercell with 36 atoms, while 'Vaca' refers to the Ti supercell with 35 atoms due to 1 vacancy. The numbers 0.1, 0.2, and 0.5 represent the maximum atomic displacements in Å within a configuration. Black circles illustrate example of chosen 30 configurations by the k-means algorithm. There is clearly visible separation into 4 groups, with Standard 0.1, 0.2, Vaca 0.1, and 0.2. However, Standard

0.5 and Vaca 0.5 have a large overlap, indicating possible similarity of their atomic environments, which is effectively taken into account with the k-means clustering.

## 3.2 Comparison of EAM Fit Quality Between K-means Clustering and Random Selection

To evaluate the effectiveness of selecting configurations via k-means clustering of their fingerprints, we compared the quality of EAM fitted on different configurations sizes obtained using this method to those obtained through uniformly random selection in terms of fit quality of energies and forces. We analyzed configuration subsets of sizes 5, 10, 15, 20, 25, 30, 40, 50, 60, 70, 80, 90, and 100, drawn from the initial total set of 1800 generated configurations. For each subset size and for each method (uniform random selection and k-means clustering), we generated 50 distinct groups of configurations. The variation between the groups in random selection came from the inherent randomness of the process, while in k-means clustering, variation resulted from the initial random positions of the cluster centroids. In the case of selecting 5 structures using k-means, some of the 50 groups contained identical structures, as the centroids converged to the same positions determined by the k-means algorithm.

Each EAM fit was evaluated by the RMSE between the fitted and DFT data for energy and forces of a test set comprising all 1800 configurations. Performing 50 fits (one for each of the 50 groups of configurations) for each subset size allowed us to gather reasonably large statistics of the fit quality for both selection methods. The comparison between the random selection and the selection by k-means clustering was then done utilizing statistical metrics of the median and the standard deviation (STD), giving two metrics for energy and two metrics for forces.

The results are summarized in **Fig. 5** for the RMSE statistics of energy and in **Fig. 6** for RMSE statistics of force. In both cases, it is evident that the k-means algorithm systematically produces better precision (better quality of the fitted EAM) with the DFT data than the uniform random selection for a smaller number of configurations. Additionally, the k-means method consistently yields very low standard deviations, unlike random selection, which exhibits higher standard deviations by a few orders up to 40 configurations. This indicates that certain fits from random selections, despite having low objective functions, significantly failed to describe the entire dataset.

In case of energy, the RMSE energy median for the k-means algorithm gradually decreases from ~0.0170 eV/atom at 5 configurations to ~0.0117 eV/atom at 20 configurations, after which it begins to oscillate. The RMSE STD of energy is larger (~0.148 eV/atom) only for 5 configurations but is close to zero (smaller than 0.01 eV/atom) for all other subset sizes, indicating that no groups of structures failed to describe the test set within its EAM fit. In contrast, with uniform random selection, the energy median also gradually decreases with a larger number of selected configurations but remains notably higher, starting at ~0.0198 eV/atom for 5 configurations, decreasing to ~0.0136 eV/atom at 20 configurations, until it meets with the k-means clustering at ~0.0120 eV/atom for 75 configurations. The STD corresponding to random selection is also larger by several orders of magnitude compared to k-means up to 30 configurations. Beyond 40 configurations, it decreases close to zero but still remains almost an order of magnitude larger than the STD for the k-means.

In case of forces, the RMSE force median for the k-means algorithm decreases from ~0.2935 eV/Å at 5 configurations to ~0.2562 eV/Å at 30 configurations, after which it starts to oscillate. With random selection, the force median gradually decreases from ~0.3144 eV/Å at 5 configurations to ~0.2708 eV/Å at 30 configurations but does not achieve the same precision as the k-means method, even at 100 configurations, where it has a median of 0.2613 eV/Å. The STD for forces follows the same pattern and differences between k-means and random selection as observed for the energies.

Considering both the RMSE for energy and forces, the fitted energies and forces demonstrate precision on the order of ~0.01 eV/atom and ~0.25 eV/Å, respectively, even when using a full set of 1800 for the fitting. This level of accuracy is reasonable for the EAM potential and is comparable to the precision achieved by, e.g., the NiTi potential fitted using the same MEAMfit2 software [20]. The use of k-means clustering to select representative configurations indicates that approximately 30 structures are sufficient to adequately describe the 1800 generated configurations in terms of energies and forces within the EAM potential framework. This approach illustrates the potential significant reduction for computational time for the extensive DFT calculations and the time required for the potential fitting.

The energy converges to the same values between the two selection methods, but the forces are consistently better with k-means clustering up to 100 configurations. This indicates larger variation in the forces among the configurations and k-means clustering, by selecting diverse and non-redundant atomic environments, captures this variation more effectively compared to the random selection.

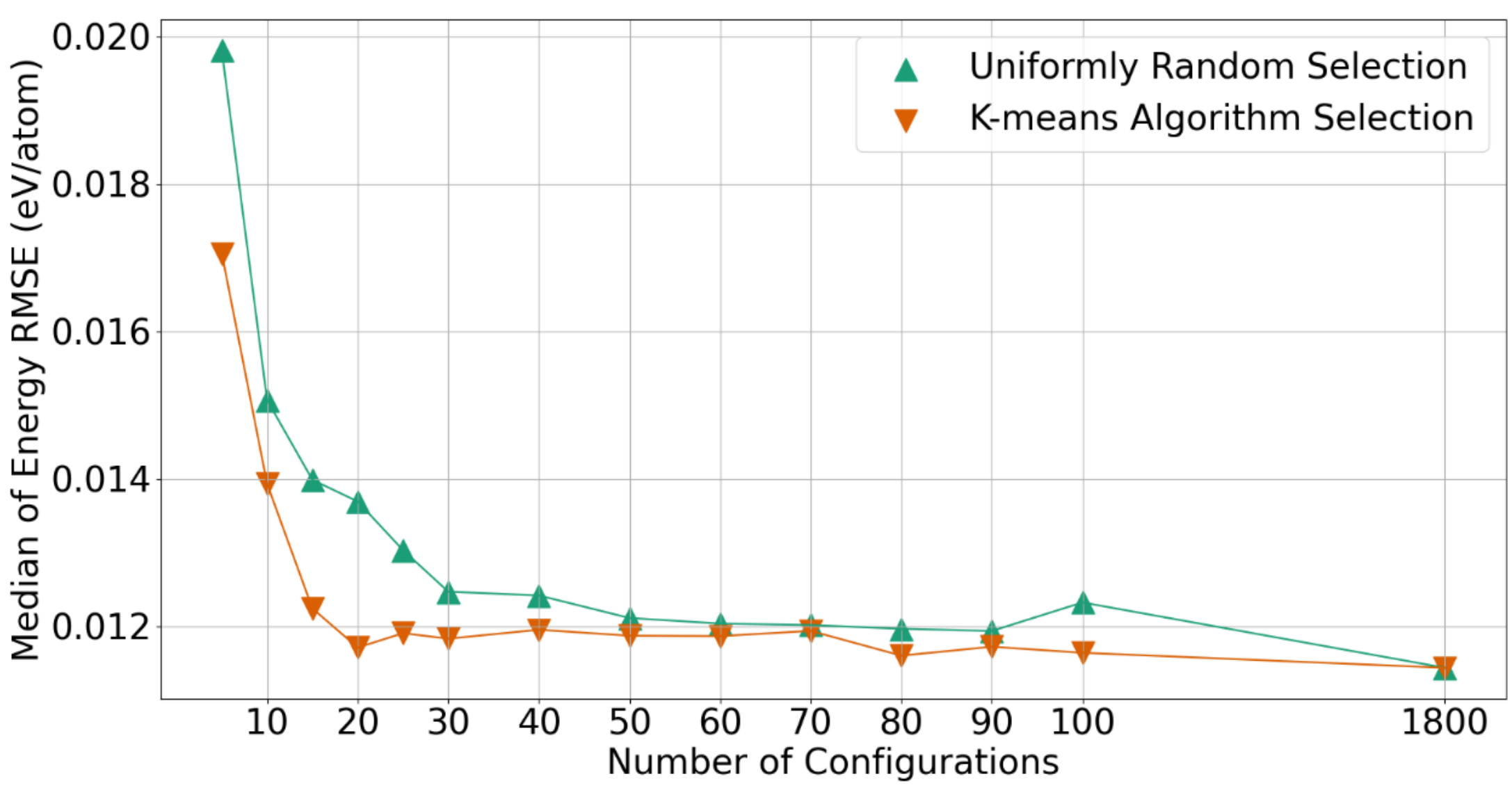

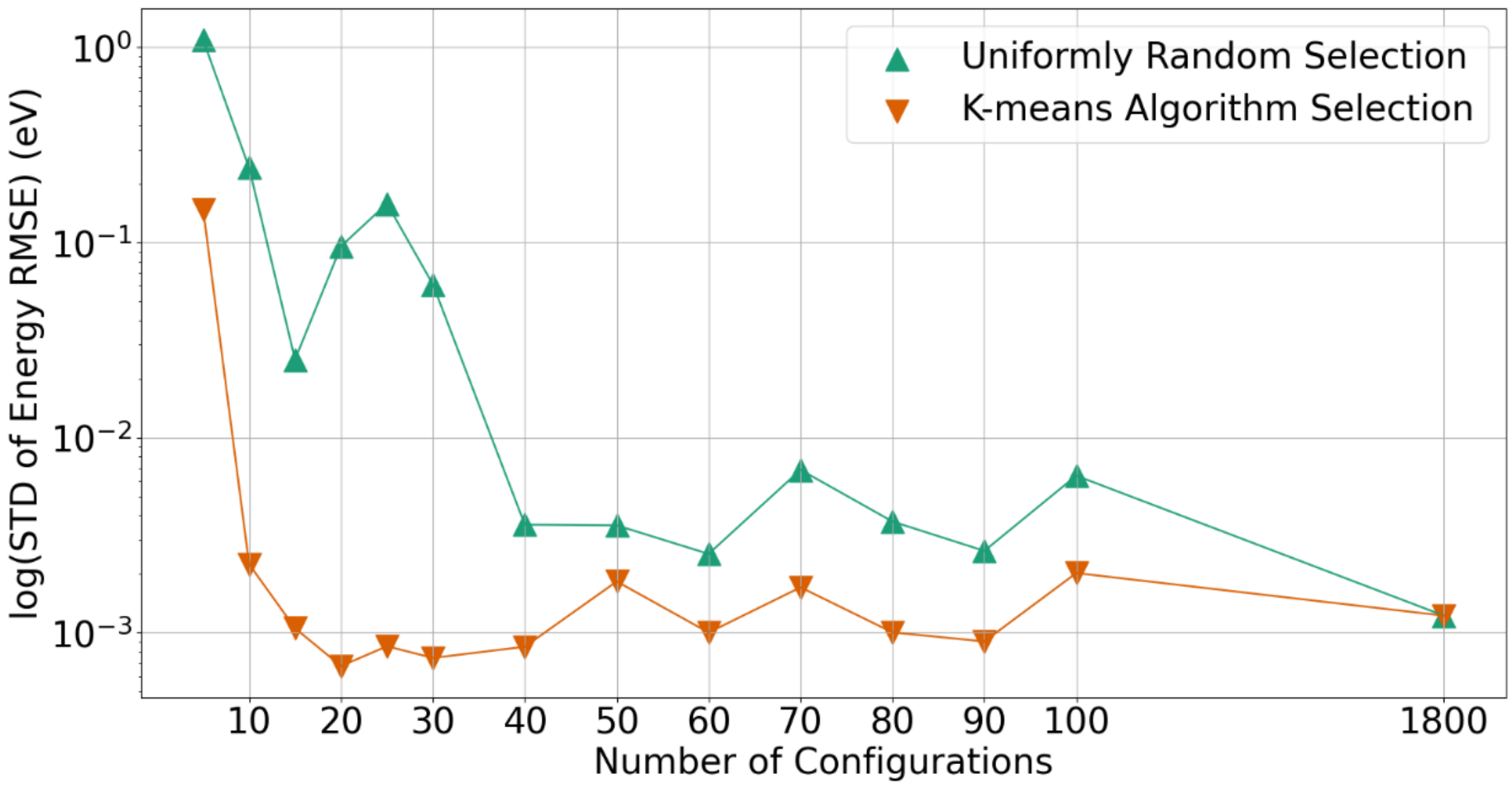

**Fig. 5**: Median and standard deviation (STD) of RMSE for energy, assessing the quality of EAM fit across different configuration sizes selected by the k-means clustering algorithm and by the uniformly random sampling. The k-means selection consistently shows better precision and lower standard deviation for smaller number of configurations compared to random selection.

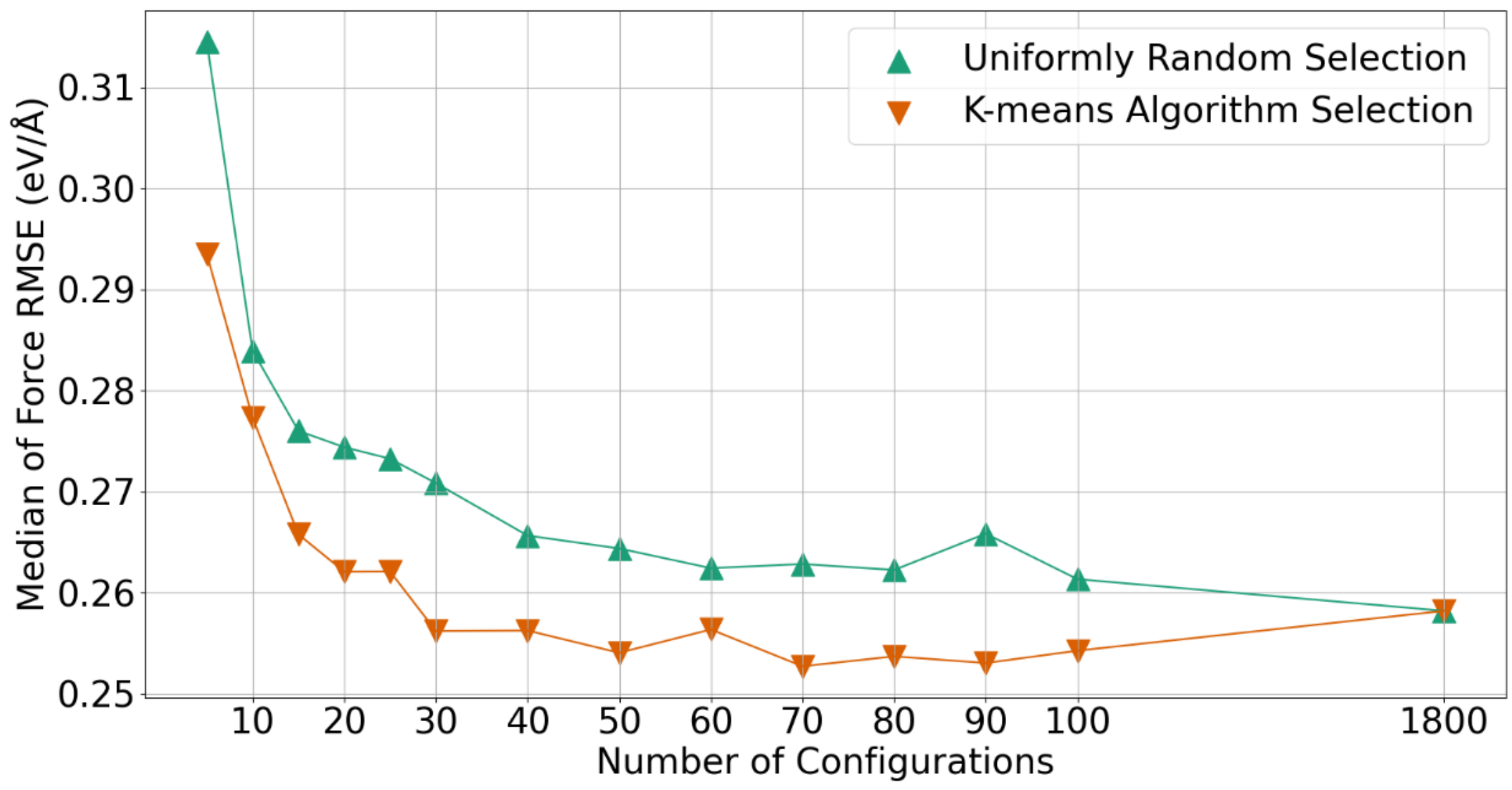

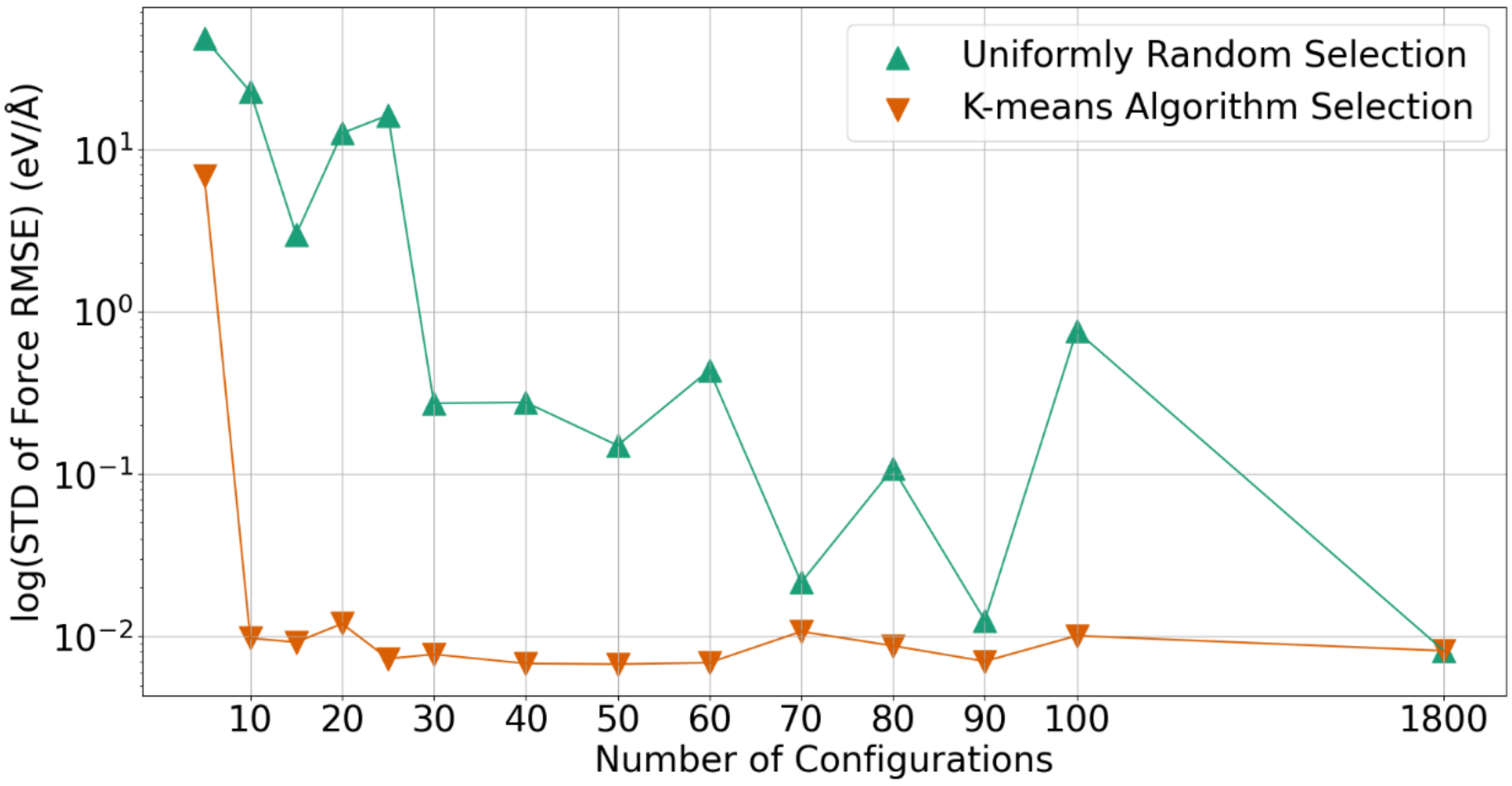

**Fig. 6**: Median and standard deviation of RMSE for forces, assessing the quality of EAM fit across different configuration sizes selected by the k-means clustering algorithm and by the uniformly random sampling. The k-means selection consistently shows better precision and lower standard deviation for smaller number of configurations compared to random selection.

### 3.3 Assessment of Predictive Power of Overlapping Configurations in t-SNE Space

In Section 3.1 in the **Fig. 3,** we showed that the t-SNE method reveals a notable overlap between configurations with vacancies (Vaca 0.5) and those without vacancies (Standard 0.5), both having a maximum displacement of 0.5 Å. This overlap suggests that the potential derived from structures without vacancies could effectively predict the energies and forces of structures with vacancies, and vice versa, due to their similar atomic environments. To test it, we removed the Vaca 0.5 configurations from the structures selected by the k-means algorithm, i.e., the structures were now selected from 1500 atomistic configurations instead of 1800. We then used k-means to select 40 configurations, repeated the selection and fitting 50 times, and used 300 Vaca 0.5 configurations as the test dataset. As summarized in **Tab. 1**, the RMSE for energy was 0.0173 eV/atom, and for forces, it was 0.3725 eV/Å. When the 40 selected configurations included some Vaca 0.5 configurations and the same 300 configurations were used as the test set, the RMSE values remained similar: The RMSE for energy slightly improved to 0.0167 eV/atom (3.5% improvement), while the RMSE for forces increased slightly to 0.3747 eV/Å (0.6% increase). In contrast, when both the Vaca 0.5 and Standard 0.5 subsets were excluded from the 40-configuration selection, the RMSE for the Vaca 0.5 test set showed significantly reduced accuracy, with values more than twice as high: 0.0380 eV/atom for energy and 0.7533 eV/Å for forces. Thus, these findings demonstrate that the overlap observed in the t-SNE 2D map indicates potential redundancy in the information (in terms of total energies and forces) between overlapping sets of configurations in the 2D space.

**Tab. 1:** Median RMSE for energy and forces on a test set of 300 Vaca 0.5 configurations, evaluating the quality of the EAM fit for the 40 configurations selected using the k-means algorithm under three scenarios: (1) no configurations were removed from the initial set of 1800 configurations, (2) Vaca 0.5 was removed from the selection pool, leaving 1500 configurations, and (3) both Vaca 0.5 and Standard 0.5 were removed from the selection pool, leaving 1200 configurations.

| **Configuration Scenario** | **RMSE of Energy (eV/atom)** | **RMSE of Forces (eV/Å)** |
|---|---|---|
| 40 configurations, no group excluded | 0.0167 | 0.3747 |
| 40 configurations, excluded Vaca 0.5 | 0.0173 | 0.3725 |
| 40 configurations, excluded Vaca 0.5 and Standard 0.5 | 0.0380 | 0.7533 |

## 4. Conclusions

This study demonstrates the effectiveness of using k-means clustering based on atomistic configuration fingerprints combining CrystalNN and RDF to select configurations from a larger generated set for fitting MD interatomic potentials on DFT data. The advantages of this selection method were shown in the fitting of the EAM potential for Ti with an initial configuration size of 1800. The results indicated that k-means clustering consistently outperforms uniform random selection, achieving higher precision with fewer configurations in terms of fitting energies and forces. Additionally, the k-means method produced lower standard deviations, indicating more reliable fits. The results also suggest that only 30 configurations selected by the k-means algorithm are necessary to achieve a reasonably precise EAM fit which describes well all 1800 configurations in terms of energies and forces.

Additionally, using configuration fingerprints reduction into 2D space by t-SNE method, a significant overlap was revealed between 300 configurations without Ti vacancy and 300 configurations with a vacancy, both subsets having a maximum atomic displacement of 0.5 Å. This overlap indicates possible redundant information (due to similarity in atomic environments) within these two subsets. This redundancy is implicitly accounted for by the k-means selection but cannot be captured by a random selection, highlighting the advantage of using the clustering method. Additionally, when configurations with vacancies were excluded from the k-means algorithm and used solely as a test set, their energy and force predictions maintained similar precision to when they were included. This finding confirms that the overlapping clusters in the 2D t-SNE space imply potentially redundant information between these configurations in terms of energies and forces.

## 5. Acknowledgments

This work was supported by the Grant Agency of the Czech Technical University in Prague [grant No. SGS24/121/OHK2/3T/12]. Computational resources were provided by the e-INFRA CZ project (ID:90254), supported by the Ministry of Education, Youth and Sports of the Czech Republic. I would also like to thank Pavel Baláž for his valuable advice, which greatly contributed to the improvement of this work.

## Table of Contents Graphic

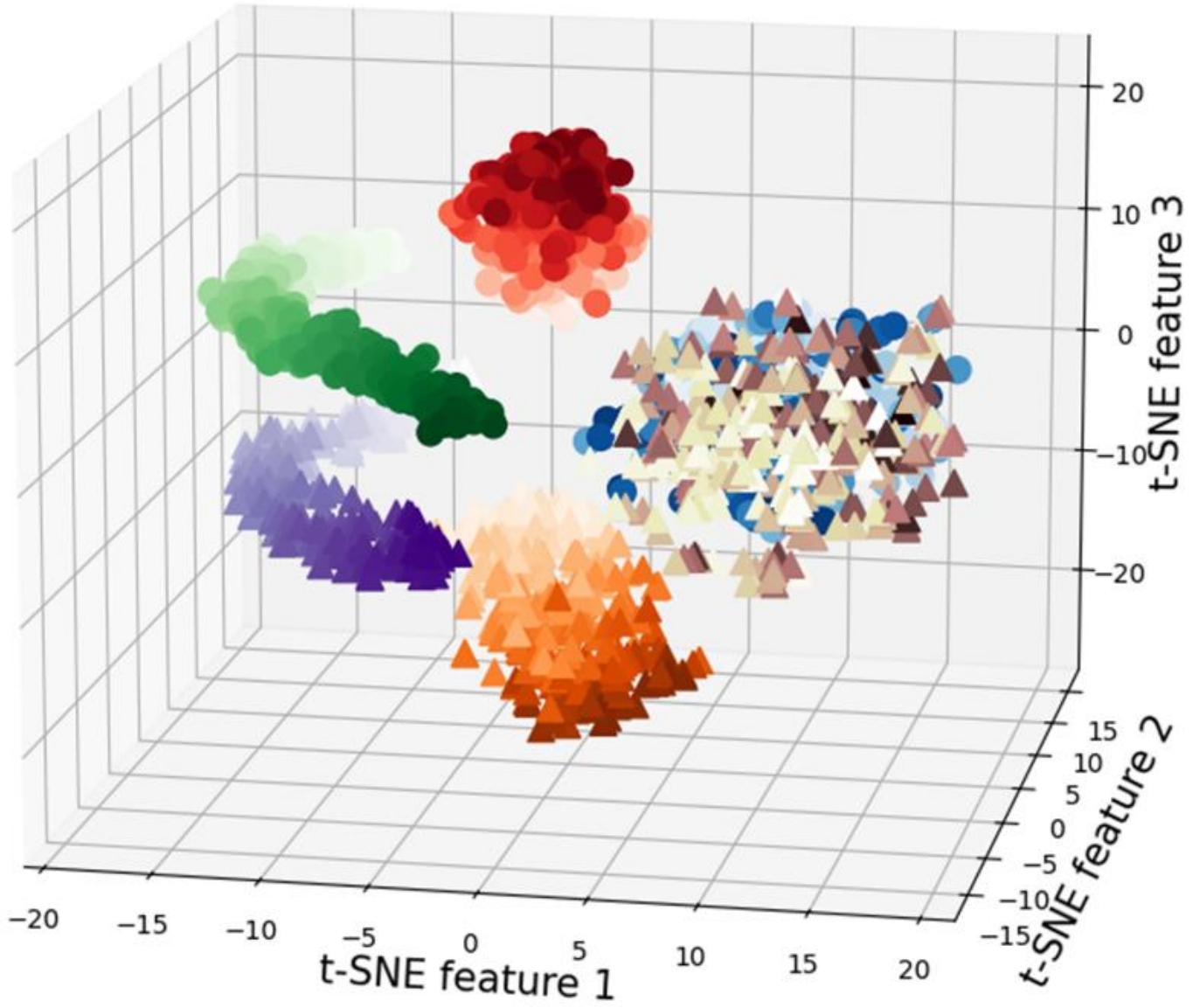